\documentclass[aps,twocolumn]{revtex4}
\usepackage{amsmath}
\usepackage{epsfig}
\usepackage{multirow}
\usepackage{appendix}
\usepackage{amssymb}
\usepackage{epstopdf}

\begin{document}
	
\title{Investigating $\Omega \phi$ Interaction and Correlation Functions}
\author{Ye Yan$^{1}$}
\author{Yuheng Wu$^{2}$}
\author{Yue Tan$^{2}$}
\author{Qi Huang$^3$}\email{06289@njnu.edu.cn(Corresponding author)}
\author{Hongxia Huang$^3$}\email{hxhuang@njnu.edu.cn(Corresponding author)}
\author{Jialun Ping$^3$}

\affiliation{$^1$Department of Physics, Changzhou College of Information Technology, Changzhou 213164, China}
\affiliation{$^2$Department of Physics, Yancheng Institute of Technology, Yancheng 224000, China}
\affiliation{$^3$Department of Physics, Nanjing Normal University, Nanjing 210023, China}

\begin{abstract}
	
In this work, we investigate the interaction between the $\Omega$ baryon and the $s\bar{s}$ meson within the framework of the quark delocalization color screening model.
The spectra calculations show that no bound state is formed in any of the considered channels, while the scattering indicates that the $\Omega\phi$ interaction with $J^{P}=1/2^{-}$ is weakly attractive. 
As for the $\Omega\phi$ interactions with $J^{P}=3/2^{-}$ and $5/2^{-}$, as well as the $\Omega\eta^{\prime}$ interaction with $J^{P}=3/2^{-}$, they are all repulsive.
After an investigation on the femtoscopic correlation functions, we find that, due to the spin-averaging effect, the overall $\Omega\phi$ correlation function exhibits a weak dependence on the source size, which provides a crucial significance of our model for future experimental examinations in relativistic heavy-ion collisions.

\end{abstract}
	

\maketitle

\setcounter{totalnumber}{5}

\section{Introduction}

Decoding the interaction between hadrons is one of the central goals of hadronic and nuclear physics.
At low energies, hadron–hadron interactions encode essential information on nonperturbative quantum chromodynamics (QCD), including color confinement, quark exchange effects, and the emergence of effective forces between color-singlet objects.
In particular, systems involving strange and multi-strange hadrons provide a unique testing ground for theoretical models, as they are less affected by light-quark exchange and Pauli blocking at the hadronic level, and are therefore expected to exhibit relatively clean manifestations of quark-level dynamics.

In recent years, growing attention has been paid to baryon--baryon and baryon--meson interactions involving multi-strange baryons, such as the $\Omega$ baryon.
The $\Omega$ baryon, being composed solely of strange quarks, plays a special role in hadron physics.
Its interactions are free from valence light-quark contributions and are thus particularly sensitive to short-range dynamics and gluon exchange.
From the experimental side, relativistic heavy-ion collisions have opened new opportunities to access $\Omega$-related interactions through femtoscopic correlation measurements, thereby compensating for the lack of experimental information from conventional scattering measurements of unstable hadrons.

Two representative examples of $\Omega$-baryon interactions that have attracted considerable attention are the $N \Omega$ with $J^{P} = 2^{+}$ and the $\Omega \Omega$ with $J^{P} = 0^{+}$.
Following its initial proposal, the $N\Omega$ system with $J^{P}=2^{+}$ has been extensively studied in various theoretical approaches.
Early quark-model calculations suggested a narrow resonance or quasi-bound state in this channel~\cite{Goldman:1987ma,Oka:1988yq}.
Later, lattice QCD simulations by the HAL QCD Collaboration demonstrated that the $N\Omega$ interaction supports a bound state, both at heavy pion masses and near the physical point~\cite{HALQCD:2014okw,HALQCD:2018qyu}.
Femtoscopic analyses based on lattice-derived interactions, together with direct correlation measurements by the STAR and ALICE Collaborations in heavy-ion and $pp$ collisions, provided experimental support for an attractive interaction~\cite{Morita:2016auo,Morita:2019rph,STAR:2018uho,ALICE:2020mfd}.
Consistent conclusions have also been obtained in other approaches, including the chromomagnetic model, QCD sum rules, and constituent quark models~\cite{Silvestre-Brac:1992xsl,Chen:2021hxs,Li:1999bc,Dai:2007gc,Huang:2015yza}.

As another prominent example, the $\Omega\Omega$ system with $J^{P}=0^{+}$ was first suggested as a possible dibaryon configuration in a chiral soliton model with explicit scalar mesons~\cite{Kopeliovich:1990pp}.
Subsequent studies based on quark models further explored its energy spectrum and stability, providing additional support for the existence of an attractive interaction in this channel\cite{Li:2000cb,Liu:2002vi}.
A quantitative investigation was later carried out by the HAL QCD Collaboration, which employed lattice QCD simulations to extract the $\Omega\Omega$ interaction and revealed an intermediate-range attraction in the $^{1}S_{0}$ channel~\cite{HALQCD:2015qmg}.
Based on the lattice-derived potential, the femtoscopic correlation function of the $\Omega\Omega$ pair was subsequently analyzed, and characteristic correlation signatures were proposed to guide future experimental searches in relativistic heavy-ion collisions~\cite{Morita:2019rph}.

Compared with the extensively studied $\Omega$–baryon interactions, investigations of $\Omega$–meson interactions remain rather limited.
The observation of the $\Omega(2012)$ resonance has renewed interest in this sector, as its internal structure is difficult to accommodate within a conventional three-quark picture.
In studies interpreting $\Omega(2012)$ as a hadronic molecular state, the $\Omega \eta$ interaction has attracted particular attention and has been explored within various theoretical frameworks, including quark models and effective field theory approaches, with the aim of elucidating the role of coupled-channel dynamics in the formation of the $\Omega(2012)$~\cite{Valderrama:2018bmv,Huang:2018wth,Pavao:2018xub,Zeng:2020och,Gutsche:2019eoh,Lu:2020ste,Ikeno:2020vqv,Liu:2020yen,Hu:2022pae,Ikeno:2022jpe,Ikeno:2023wyh,Song:2024ejc,Xie:2024wbd}.
In addition, the interaction between the $\Omega$ baryon and charmonium was investigated within a potential quark model in Ref.~\cite{Meng:2019fan}.

In recent years, significant progress has been achieved in both experimental measurements~\cite{ALICE:2023eyl,ALICE:2023zbh,Hu:2023iib,ALICE:2025wuy,Fu:2024btw,ALICE:2025zzg,STAR:2024zvj,ALICE:2025plu,STAR:2025jwe,ALICE:2020mfd,ALICE:2021cpv} and theoretical studies~\cite{Encarnacion:2025luc,Etminan:2024nak,Encarnacion:2025lyf,Li:2024tvo,Jinno:2024tjh,Agatao:2025ckp,Ikeno:2025bsx,Encarnacion:2024jge,Ge:2025put,Liu:2025eqw,Tang:2025bcc,Liu:2025oar,Zeng:2025kur,Wang:2025kcy,Liu:2025nze,Zhang:2025tfd,Lin:2025mtz,Torres-Rincon:2024znb} of femtoscopic correlation functions, driven by advances in detection techniques and data analysis methods.
In particular, attractive strong interactions between the $\Omega$ baryon and the proton~\cite{ALICE:2020mfd}, as well as between the $\phi$ meson and the proton~\cite{ALICE:2021cpv}, have been experimentally supported by the ALICE Collaboration using femtoscopic techniques.
These observations naturally motivate further investigations of the $\Omega \phi$ interaction and the corresponding femtoscopic correlation functions, which may provide new insights into hadron–hadron interactions involving multi-strange systems.
Moreover, due to the Pauli principle acting at the quark level, the total spin configuration of the system can strongly influence the effective interaction, leading to different behaviors in different spin-parity channels.

Motivated by these developments, in this work we perform a systematic study of the $\Omega$--$s\bar{s}$ meson interaction, focusing on the $\Omega \phi$ and $\Omega \eta^{\prime}$ systems in the $S$-wave.
We first investigate the possible existence of bound state, analyze the low-energy scattering behavior in different spin-parity channels, and extract the corresponding scattering parameters.
Based on these results, we further evaluate the femtoscopic correlation functions using the Lednický-Lyuboshits (LL) model, aiming to provide theoretical guidance for future experimental studies of $\Omega$–meson interactions.
In our previous work, the $p\Omega$ interaction and its femtoscopic correlation functions were studied within the quark delocalization color screening model (QDCSM)~\cite{Yan:2024aap}.
The QDCSM is a constituent quark model incorporating quark delocalization and color screening, which are responsible for generating intermediate-range attraction in baryon–baryon interactions~\cite{Wang:1992wi,Ping:1998si,Wu:1998wu}.
It has been successfully applied to various dibaryon and pentaquark systems, including the $d^{*}$, $N \Omega$, $P_c$ and $sscq\bar{q}$ states~\cite{Ping:2008tp,Huang:2015uda,Yan:2023tvl,Yan:2024aap}.
On this basis, we extend the QDCSM to investigate the $\Omega$--$s\bar{s}$ interaction in the present work.

The structure of this paper is as follows: 
The detail of the QDCSM are presented in the next section. 
In Sec.~\ref{3}, numerical analysis of the $\Omega$--$s\bar{s}$ systems is carried out, including the energy spectrum, scattering phase shifts, and correlation functions.
Finally, Sec.~\ref{4} summarizes the main conclusions of this work.

\section{THEORETICAL FORMALISM}
\label{2}

\subsection{Quark delocalization color screening model}
\label{21}

In this section, we briefly introduce the salient features of the used quark model.
The QDCSM is an extension of the quark cluster model~\cite{Wang:1992wi} and the general form of the pentaquark Hamiltonian is given by:
	\begin{align}
	H=&\sum_{i=1}^5\left(m_i+\frac{\boldsymbol{p}_{i}^{2}}{2m_i}\right)-T_{\mathrm{c} . \mathrm{m}} +\sum_{j>i=1}^5 V(\boldsymbol{r}_{ij}),
\end{align}
where $m_i$ is the quark mass, $\boldsymbol{p}_{i}$ is the momentum of the quark, and $T_{\mathrm{c.m.}}$ is the center-of-mass kinetic energy.
The dynamics of the hexaquark system is driven by two-body potentials, including color confinement ($V_{\mathrm{CON}}$), perturbative one-gluon exchange interaction ($V_{\mathrm{OGE}}$), and dynamical chiral symmetry breaking ($V_{\chi}$).
\begin{align}
	V(\boldsymbol{r}_{ij})= & V_{\mathrm{CON}}(\boldsymbol{r}_{ij})+V_{\mathrm{OGE}}(\boldsymbol{r}_{ij})+V_{\chi}(\boldsymbol{r}_{ij}).
\end{align}

Here, a phenomenological color screening confinement potential ($V_{\mathrm{CON}}$) is used as:
\begin{align}
	V_{\mathrm{CON}}(\boldsymbol{r}_{ij}) = & -a_{c}\boldsymbol{\lambda}_{i}^{c} \cdot \boldsymbol{\lambda}_{j}^{c}\left[  f(\boldsymbol{r}_{ij})+V_{0}\right],
\end{align}
\begin{align}
	f(\boldsymbol{r}_{ij}) =& \left\{\begin{array}{l}
		\boldsymbol{r}_{i j}^{2}, ~~~~~~~~~~~~~ ~i,j ~\text {occur in the same cluster } \\
		\frac{1-e^{-\mu_{q_{i}q_{j}} \boldsymbol{r}_{i j}^{2}}}{\mu_{q_{i}q_{j}}},  ~~~i,j ~\text {occur in different cluster }
	\end{array}\right.   \nonumber
\end{align}
where $a_c$, $V_{0}$ and $\mu_{q_{i}q_{j}}$ are model parameters, and $\boldsymbol{\lambda}^{c}$ stands for the SU(3) color Gell-Mann matrices.
Among them, the color screening parameter $\mu_{q_{i}q_{j}}$ is determined by fitting the deuteron properties, nucleon-nucleon scattering phase shifts, and hyperon-nucleon scattering phase shifts, respectively, with $\mu_{qq}=0.45$, $\mu_{qs}=0.19$, and $\mu_{ss}=0.08~$fm$^{-2}$, satisfying the relation $\mu_{qs}^{2}=\mu_{qq}\mu_{ss}$~\cite{Chen:2011zzb}.

In the present work, we mainly focus on the low-lying negative parity system of the $S$-wave, so the spin-orbit and tensor interactions are not included.
The one-gluon exchange potential ($V_{\mathrm{OGE}}$), which includes Coulomb and chromomagnetic interactions, is written as:
\begin{align}
	V_{\mathrm{OGE}}(\boldsymbol{r}_{ij})= &\frac{1}{4}\alpha_{s} \boldsymbol{\lambda}_{i}^{c} \cdot \boldsymbol{\lambda}_{j}^{c}  \\
	&\cdot \left[\frac{1}{r_{i j}}-\frac{\pi}{2} \delta\left(\mathbf{r}_{i j}\right)\left(\frac{1}{m_{i}^{2}}+\frac{1}{m_{j}^{2}}+\frac{4 \boldsymbol{\sigma}_{i} \cdot \boldsymbol{\sigma}_{j}}{3 m_{i} m_{j}}\right)\right],   \nonumber \label{Voge}
\end{align}
where $\boldsymbol{\sigma}$ is the Pauli matrices and $\alpha_{s}$ is the quark-gluon coupling constant.

The dynamical breaking of chiral symmetry results in the SU(3) Goldstone boson exchange interactions appear between constituent light quarks $u, d$, and $s$.
Hence, the chiral interaction is expressed as:
\begin{align}
	V_{\chi}(\boldsymbol{r}_{ij})= & V_{\pi}(\boldsymbol{r}_{ij})+V_{K}(\boldsymbol{r}_{ij})+V_{\eta}(\boldsymbol{r}_{ij}).
\end{align}
Among them:
\begin{align}
	V_{\pi}\left(\boldsymbol{r}_{i j}\right) =&\frac{g_{c h}^{2}}{4 \pi} \frac{m_{\pi}^{2}}{12 m_{i} m_{j}} \frac{\Lambda_{\pi}^{2}}{\Lambda_{\pi}^{2}-m_{\pi}^{2}} m_{\pi}\left[Y\left(m_{\pi} \boldsymbol{r}_{i j}\right)\right. \nonumber \\
	&\left.-\frac{\Lambda_{\pi}^{3}}{m_{\pi}^{3}} Y\left(\Lambda_{\pi} \boldsymbol{r}_{i j}\right)\right]\left(\boldsymbol{\sigma}_{i} \cdot \boldsymbol{\sigma}_{j}\right) \sum_{a=1}^{3}\left(\boldsymbol{\lambda}_{i}^{a} \cdot \boldsymbol{\lambda}_{j}^{a}\right),
\end{align}
\begin{align}
	V_{K}\left(\boldsymbol{r}_{i j}\right) =&\frac{g_{c h}^{2}}{4 \pi} \frac{m_{K}^{2}}{12 m_{i} m_{j}} \frac{\Lambda_{K}^{2}}{\Lambda_{K}^{2}-m_{K}^{2}} m_{K}\left[Y\left(m_{K} \boldsymbol{r}_{i j}\right)\right. \nonumber \\
	&\left.-\frac{\Lambda_{K}^{3}}{m_{K}^{3}} Y\left(\Lambda_{K} \boldsymbol{r}_{i j}\right)\right]\left(\boldsymbol{\sigma}_{i} \cdot \boldsymbol{\sigma}_{j}\right) \sum_{a=4}^{7}\left(\boldsymbol{\lambda}_{i}^{a} \cdot \boldsymbol{\lambda}_{j}^{a}\right),
\end{align}
\begin{align}
	V_{\eta}\left(\boldsymbol{r}_{i j}\right) =&\frac{g_{c h}^{2}}{4 \pi} \frac{m_{\eta}^{2}}{12 m_{i} m_{j}} \frac{\Lambda_{\eta}^{2}}{\Lambda_{\eta}^{2}-m_{\eta}^{2}} m_{\eta}\left[Y\left(m_{\eta} \boldsymbol{r}_{i j}\right)\right. \nonumber \\
	&\left.-\frac{\Lambda_{\eta}^{3}}{m_{\eta}^{3}} Y\left(\Lambda_{\eta} \boldsymbol{r}_{i j}\right)\right]\left(\boldsymbol{\sigma}_{i} \cdot \boldsymbol{\sigma}_{j}\right)\left[\cos \theta_{p}\left(\boldsymbol{\lambda}_{i}^{8} \cdot \boldsymbol{\lambda}_{j}^{8}\right)\right.  \nonumber \\
	&\left.-\sin \theta_{p}\left(\boldsymbol{\lambda}_{i}^{0} \cdot \boldsymbol{\lambda}_{j}^{0}\right)\right],
\end{align}
where $Y(x) = e^{-x}/x$ is the standard Yukawa function.
The physical $\eta$ meson is considered by introducing the angle $\theta_{p}$ instead of the octet one.
 The $\boldsymbol{\lambda}^a$ are the SU(3) flavor Gell-Mann matrices.
The values of $m_\pi$, $m_K$, and $m_\eta$ are the masses of the SU(3) Goldstone bosons, which adopt the experimental values~\cite{Workman:2022ynf}.
For the $ssss\bar{s}$ pentaquark system, the $\pi$- and $K$-exchange interactions vanish due to the absence of light quarks. 
Only the $\eta$-exchange interaction survives formally; however, its contribution is suppressed because of the large strange-quark mass and the short-range nature of the interaction. 
As a result, the dynamics of the system is mainly governed by the color interactions, including the one-gluon-exchange and confinement potentials.

In the present work, the same chiral coupling constant $g_{ch}$ as in the light-quark sector is adopted for consistency, which is determined from the $\pi N N$ coupling constant through:
\begin{align}
	\frac{g_{c h}^{2}}{4 \pi} & = \left(\frac{3}{5}\right)^{2} \frac{g_{\pi N N}^{2}}{4 \pi} \frac{m_{u, d}^{2}}{m_{N}^{2}}.
\end{align}
The other symbols in the above expressions have their usual meanings.
The parameters used in this work are the same as those adopted in our previous studies of the interpretation of excited $\Omega_c$ from a pentaquark perspective~\cite{Yan:2023tvl} and the $N\Omega$ system~\cite{Huang:2015yza,Yan:2024aap}.
Table~\ref{parameters} lists the model parameters, while Table~\ref{hadrons} summarizes the calculated masses of the hadrons.

\begin{table}[ht]
	\caption{\label{parameters}Model parameters used in this work:
		$m_{\pi} = 0.7$, $m_{K} = 2.51$, $m_{\eta} = 2.77$, $\Lambda_{\pi} = 4.2$, $\Lambda_{K} = 5.2$, $\Lambda_{\eta} = 5.2$ fm$^{-1}$, $g_{ch}^2/(4\pi)$ = 0.54.}
	\begin{tabular}{cccccc} \hline\hline
		~~~~$b$~~~~ & ~~$m_{q}$~~ & ~~~$m_{s}$~~~  & ~~~$V_{0_{qq}}$~~~~&~~~$V_{0_{q\bar{q}}}$~~~~& ~~~$ a_c$~~~   \\
		(fm)        & (MeV)       & (MeV)          & (fm$^{-2}$)        & (fm$^{-2}$)             & ~(MeV\,fm$^{-2}$)~ \\
		0.518       & 313         & 573            &  $-$1.288          & $-$0.743                &  58.03 \\ \hline
		$\alpha_{s_{\bar{q}q}}$ &$\alpha_{s_{\bar{s}q}}$  & $\alpha_{s_{\bar{s}s}}$ &$\alpha_{s_{qq}}$   & $\alpha_{s_{qs}}$ &$\alpha_{s_{ss}}$   \\
		1.491   & 1.427     &  1.004  &    0.565                  & 0.524             &   0.451         \\ \hline\hline	
	\end{tabular}
\end{table}

\begin{table}[ht]
	\caption{The masses (in MeV) of the baryons and mesons. Experimental values are taken from the Particle Data Group~\cite{Workman:2022ynf}.}
	\begin{tabular}{c c c |c c c}
		\hline \hline
		~Baryon~ & ~$M^{\text{Exp}}$~ & ~$M^{\text{Theo}}$~ & ~Meson~ & ~$M^{\text{Exp}}$~ & ~$M^{\text{Theo}}$~ \\ \hline
		$N$    & 939   & 939  & $\pi$ & 139 & 139 \\
		$\Delta$ & 1232  & 1232 & $\rho$ & 770 & 770 \\
		$\Lambda$ & 1115  & 1123 & $\eta$ & 582 & 283 \\
		$\Sigma$  & 1189  & 1238 & $\omega$ & 782 & 721 \\
		$\Sigma^*$ & 1385  & 1360  &  $K$ & 495  & 495 \\
		$\Xi$ & 1318  & 1375 & $K^*$ & 892 & 814 \\
		$\Xi^*$ & 1536  & 1496 &  $\eta^\prime$ & 958 & 912 \\
		$\Omega$ & 1672  & 1642 &  $\phi$ & 1020 & 1020 \\  
		
		\hline\hline
	\end{tabular}
	\label{hadrons}
\end{table}

In the QDCSM, quark delocalization was introduced to enlarge the model variational space to take into account the mutual distortion or the internal excitations of nucleons in the course of interaction.
It is realized by specifying the single-particle orbital wave function of the QDCSM as a linear combination of left and right Gaussians, the single-particle orbital wave functions used in the ordinary quark cluster model
\begin{eqnarray}
	\psi_{\alpha}(\boldsymbol {S_{i}} ,\epsilon) & = & \left(
	\phi_{\alpha}(\boldsymbol {S_{i}})
	+ \epsilon \phi_{\alpha}(-\boldsymbol {S_{i}})\right) /N(\epsilon), \nonumber \\
	\psi_{\beta}(-\boldsymbol {S_{i}} ,\epsilon) & = &
	\left(\phi_{\beta}(-\boldsymbol {S_{i}})
	+ \epsilon \phi_{\beta}(\boldsymbol {S_{i}})\right) /N(\epsilon), \nonumber \\
	N(S_{i},\epsilon) & = & \sqrt{1+\epsilon^2+2\epsilon e^{-S_i^2/4b^2}}. \label{1q}
\end{eqnarray}
It is worth noting that the mixing parameter $\epsilon$ is not an adjusted one but determined variationally by the dynamics of the multiquark system itself.
In this way, the multiquark system chooses its favorable configuration in the interacting process.
This mechanism has been used to explain the crossover transition between the hadron phase and quark-gluon plasma phase~\cite{Xu:2007oam}.

\subsection{Resonating group method for bound-state and scattering process}
\label{22}

The resonating group method (RGM)~\cite{RGM1,RGM} and generating coordinates method~\cite{GCM1,GCM2} are used to carry out a dynamical calculation.
The main feature of the RGM for two-cluster systems is that it assumes that two clusters are frozen inside, and only considers the relative motion between the two clusters.
So the conventional ansatz for the two-cluster wave functions is
\begin{equation}
	\psi_{5q} = {\cal A }\left[[\phi_{B}\phi_{M}]^{[\sigma]IS}\otimes\chi(\boldsymbol{R})\right]^{J}, \label{5q}
\end{equation}
where the symbol ${\cal A }$ is the antisymmetrization operator, and ${\cal A} = 1-P_{14}-P_{24}-P_{34}$. $[\sigma]=[222]$ gives the total color symmetry and all other symbols have their usual meanings.
$\phi_{B}$ and $\phi_{M}$ are the $q^{3}$ and $\bar{q}q$ cluster wave functions, respectively.
From the variational principle, after variation with respect to the relative motion wave function $\chi(\boldsymbol{\mathbf{R}})=\sum_{L}\chi_{L}(\boldsymbol{\mathbf{R}})$, one obtains the RGM equation:
\begin{equation}
	\int H(\boldsymbol{\mathbf{R}},\boldsymbol{\mathbf{R'}})\chi(\boldsymbol{\mathbf{R'}})d\boldsymbol{\mathbf{R'}}=E\int N(\boldsymbol{\mathbf{R}},\boldsymbol{\mathbf{R'}})\chi(\boldsymbol{\mathbf{R'}})d\boldsymbol{\mathbf{R'}},  \label{RGM eq}
\end{equation}
where $H(\boldsymbol{\mathbf{R}},\boldsymbol{\mathbf{R'}})$ and $N(\boldsymbol{\mathbf{R}},\boldsymbol{\mathbf{R'}})$ are Hamiltonian and norm kernels.
By solving the RGM equation, we can get the energies $E$ and the wave functions.
In fact, it is not convenient to work with the RGM expressions.
Then, we expand the relative motion wave function $\chi(\boldsymbol{\mathbf{R}})$ by using a set of Gaussians with different centers
\begin{align}
	\chi(\boldsymbol{R}) =& \frac{1}{\sqrt{4 \pi}}\left(\frac{6}{5 \pi b^{2}}\right)^{3 / 4} \sum_{i,L,M} C_{i,L} \nonumber     \\
	&\times \int \exp \left[-\frac{3}{5 b^{2}}\left(\boldsymbol{R}-\boldsymbol{S}_{i}\right)^{2}\right] Y_{L,M}\left(\hat{\boldsymbol{S}}_{i}\right) d \Omega_{\boldsymbol{S}_{i}}
\end{align}
where $L$ is the orbital angular momentum between two clusters, and $\boldsymbol {S_{i}}$, $i=1,2,...,n$ are the generator coordinates, which are introduced to expand the relative motion wave function. By including the center-of-mass motion:
\begin{equation}
	\phi_{C} (\boldsymbol{R}_{C}) = (\frac{5}{\pi b^{2}})^{3/4}e^{-\frac{5\boldsymbol{R}^{2}_{C}}{2b^{2}}},
\end{equation}
the ansatz Eq.~(\ref{5q}) can be rewritten as
\begin{align}
	\psi_{5 q} =& \mathcal{A} \sum_{i,L} C_{i,L} \int \frac{d \Omega_{\boldsymbol{S}_{i}}}{\sqrt{4 \pi}} \prod_{\alpha=1}^{3} \phi_{\alpha}\left(\boldsymbol{S}_{i}\right) \prod_{\beta=4}^{5} \phi_{\beta}\left(-\boldsymbol{S}_{i}\right) \nonumber \\
	& \times \left[\left[\chi_{I_{1} S_{1}}\left(B\right) \chi_{I_{2} S_{2}}\left(M\right)\right]^{I S} Y_{LM}\left(\hat{\boldsymbol{S}}_{i}\right)\right]^{J} \nonumber \\
	& \times \left[\chi_{c}\left(B\right) \chi_{c}\left(M\right)\right]^{[\sigma]}, \label{5q2}
\end{align}
where $\chi_{I_{1}S_{1}}$ and $\chi_{I_{2}S_{2}}$ are the product of the flavor and spin wave functions, and $\chi_{c}$ is the color wave function.
These will be shown in detail later.
$\phi_{\alpha}(\boldsymbol{S}_{i})$ and $\phi_{\beta}(-\boldsymbol{S}_{i})$ are the single-particle orbital wave functions with different reference centers,
\begin{align}
	\phi_{\alpha}\left(\boldsymbol{S}_{i}\right) & = \left(\frac{1}{\pi b^{2}}\right)^{3 / 4} e^{-\frac{1}{2 b^{2}}\left(r_{\alpha}-\frac{2}{5} \boldsymbol{S}_{i}\right)^{2}}, \nonumber \\
	\phi_{\beta}\left(\boldsymbol{-S}_{i}\right) & = \left(\frac{1}{\pi b^{2}}\right)^{3 / 4} e^{-\frac{1}{2 b^{2}}\left(r_{\beta}+\frac{3}{5} \boldsymbol{S}_{i}\right)^{2}}.
\end{align}
With the reformulated ansatz Eq.~(\ref{5q2}), the RGM Eq.~(\ref{RGM eq}) becomes an algebraic eigenvalue equation:
\begin{equation}
	\sum_{j} C_{j}H_{i,j}= E \sum_{j} C_{j}N_{i,j},
\end{equation}
where $H_{i,j}$ and $N_{i,j}$ are the Hamiltonian matrix elements and overlaps, respectively.
By solving the generalized eigenproblem, we can obtain the energy and the corresponding wave functions of the pentaquark systems.

For a scattering problem, the relative wave function is expanded as
\begin{align}
	\chi_{L}(\mathbf{R}) & =\sum_{i} C_{i} \frac{\tilde{u}_{L}\left(\boldsymbol{R}, \boldsymbol{S}_{i}\right)}{\boldsymbol{R}} Y_{L,M}(\hat{\boldsymbol{R}}),
\end{align}
with
\begin{align}
	\tilde{u}_{L}\left(\boldsymbol{R}, \boldsymbol{S}_{i}\right) & = \left\{\begin{array}{ll}
		\alpha_{i} u_{L}\left(\boldsymbol{R}, \boldsymbol{S}_{i}\right), & \boldsymbol{R} \leq \boldsymbol{R}_{C} \\
		{\left[h_{L}^{-}(\boldsymbol{k}, \boldsymbol{R})-s_{i} h_{L}^{+}(\boldsymbol{k}, \boldsymbol{R})\right] R_{A B},} & \boldsymbol{R} \geq \boldsymbol{R}_{C}
	\end{array}\right.
\end{align}
where
\begin{align}
	u_{L}\left(\boldsymbol{R}, \boldsymbol{S}_{i}\right)= & \sqrt{4 \pi}\left(\frac{6}{5 \pi b^{2}}\right)^{3 / 4} \mathbf{R} e^{-\frac{3}{5 b^{2}}\left(\boldsymbol{R}-\boldsymbol{S}_{i}\right)^{2}} \nonumber \\
	& \times i^{L} j_{L}\left(-i \frac{6}{5 b^{2}} S_{i}\right).
\end{align}

$h^{\pm}_L$ are the $L$th spherical Hankel functions, $k$ is the momentum of the relative motion with $k=\sqrt{2 \mu E_{i e}}$, $\mu$ is the reduced mass of two hadrons of the open channel, $E_{i e}$ is the incident energy of the relevant open channels, which can be written as $E_{i e} = E_{total} - E_{th}$, where $E_{total}$ denotes the total energy, and $E_{th}$ represents the threshold of the open channel.
$R_C$ is a cutoff radius beyond which all the strong interaction can be disregarded.
Additionally, $\alpha_i$ and $s_i$ are complex parameters that are determined by the smoothness condition at $R = R_C$ and $C_i$ satisfy $\sum_i C_i = 1$. After performing the variational procedure, a $L$th partial-wave equation for the scattering problem can be deduced as
\begin{align}
	\sum_j \mathcal{L}_{i j}^L C_j &= \mathcal{M}_i^L(i=0,1, \ldots, n-1),
\end{align}
with
\begin{align}
	\mathcal{L}_{i j}^L&=\mathcal{K}_{i j}^L-\mathcal{K}_{i 0}^L-\mathcal{K}_{0 j}^L+\mathcal{K}_{00}^L, \nonumber \\
	\mathcal{M}_i^L&=\mathcal{K}_{00}^L-\mathcal{K}_{i 0}^L,
\end{align}
and
\begin{align}
	\mathcal{K}_{i j}^L= & \left\langle\hat{\phi}_A \hat{\phi}_B \frac{\tilde{u}_L\left(\boldsymbol{R}^{\prime}, \boldsymbol{S}_i\right)}{\boldsymbol{R}^{\prime}} Y_{L,M}\left(\boldsymbol{R}^{\prime}\right)|H-E|\right. \nonumber \\
	& \left.\times \mathcal{A}\left[\hat{\phi}_A \hat{\phi}_B \frac{\tilde{u}_L\left(\boldsymbol{R}, \boldsymbol{S}_j\right)}{\boldsymbol{R}} Y_{L,M}(\boldsymbol{R})\right]\right\rangle .
\end{align}
By solving Eq.~(A11), we can obtain the expansion coefficients $C_i$, then the $S$-matrix element $S_L$ and the phase shifts $\delta_L$ are given by
\begin{align}
	S_L&=e^{2 i \delta_L}=\sum_{i} C_i s_i.
\end{align}

Resonances are unstable particles usually observed as bell-shaped structures in scattering cross sections of their open channels.
For a simple narrow resonance, its fundamental properties correspond to the visible cross section features: mass $M$ is at the peak position, and decay width $\Gamma$ is the half-width of the bell shape.
The cross section $\sigma_{L}$ and the scattering phase shifts $\delta_{L}$ have relations
\begin{align}
	\sigma_L&=\frac{4 \pi}{k^2}(2 L+1) \sin ^2 \delta_L.
\end{align}
Therefore, resonances can also usually be observed in the scattering phase shift, where the phase shift of the scattering channels rises through $\pi/2$ at a resonance mass.
We can obtain a resonance mass at the position of the phase shift of $\pi/2$.
The decay width is the mass difference between the phase shift of $3\pi/4$ and $\pi/4$.

\subsection{Two-particle correlation function}
\label{23}

The correlation function can be calculated using the Koonin$-$Pratt (KP) formula~\cite{Koonin:1977fh,Pratt:1990zq,Bauer:1992ffu}:
\begin{align}
	C(\boldsymbol{k}) & = \frac{N_{12}\left(\boldsymbol{p}_{1}, \boldsymbol{p}_{2}\right)}{N_{1}\left(\boldsymbol{p}_{1}\right) N_{2}\left(\boldsymbol{p}_{2}\right)}  \simeq \int \mathrm{d} \boldsymbol{r} S_{12}(r)|\Psi(\boldsymbol{r}, \boldsymbol{k})|^{2},
	\label{ignore}
\end{align}
where  $\Psi(\boldsymbol{r}, \boldsymbol{k})$ is the relative wave function in the two-body outgoing state with an asymptotic relative momentum $\boldsymbol{k}$ and $S_{12}(r)$ is the normalized pair source function in the relative coordinate, given by the expression:
\begin{align}
	S_{12}(r) = \frac{1}{(4 \pi R^2)^{3/2}} \text{exp}(-\frac{r^2}{4R^2}),
	\label{source}
\end{align}
where $R$ is the size parameter of the source.
For a pair of non-identical particles, such as $\Omega \phi$, assuming that only $S$-wave part of the wave function is modified by the two-particle interaction, $\Psi(\boldsymbol{r}, \boldsymbol{k})$ can be given by:
\begin{align}
	\Psi_{\Omega \phi}(\boldsymbol{r}, \boldsymbol{k}) = \text{exp}(\text{i} \boldsymbol{k} \cdot \boldsymbol{r}) -j_0(kr) + \psi_{\Omega \phi}(r,k).
\end{align}
Substituting the relative wave function $\Psi_{p \bar{\Omega}}(\boldsymbol{r}, \boldsymbol{k})$ into the KP formula yields the correlation function:
\begin{align}
	C_{\Omega \phi}(k) = 1 + \int_{0}^{\infty} 4\pi r^2 \, \mathrm{d}r \, S_{12}(r) \, [ |\psi_{\Omega \phi}(r,k)|^2 - |j_0(kr)|^2 ]. \label{Ck}
\end{align}
By assuming that the relative wave function $\psi_{\Omega\phi}(r,k)$ satisfies the asymptotic scattering form, the expression of the correlation function in the LL model can be derived as:
\begin{align}
	C(k) =& 1 + \frac{|f(k)|^2}{2R^2} \left(1 - \frac{r_{\mathrm{0}}}{2\sqrt{\pi} R}\right)  \nonumber \\ 
	& + \frac{2 \text{Re} f(k)}{\sqrt{\pi} R}  \, F_1(2kR) - \frac{\text{Im}f(k)}{R} \, F_2(2kR).
\end{align}
where $f(k)$ denotes the scattering amplitude, which can be expressed in terms of the effective range expansion.
The remaining terms are given by:
\begin{align}
	f(k) &=\left[k \cot \delta(k)- \text{i} k\right]^{-1} \approx\left[-\frac{1}{a_{0}}+\frac{1}{2} r_{\text{eff}} k^{2}- \text{i} k\right]^{-1}, \\
	F_{1}(x) & =\int_{0}^{x} \frac{1}{x} \exp \left(t^{2}-x^{2}\right) \mathrm{d} t, \\
	F_{2}(x) & =\left(1-\exp \left(-x^{2}\right)\right) / x, \\
	x & =2 k R. 
\end{align}
Additionally, for the $S$-wave $\Omega \phi$ system, the possible spin-parity quantum numbers are $J^P = 1/2^-$, $3/2^-$, and $5/2^-$, respectively.
Since the experimentally measured correlation function is spin-averaged, the theoretically obtained correlation function should also consider the average over systems with different quantum numbers:
\begin{align}
	C_{\Omega \phi}^{t o t a l}(k)=\frac{1}{6} C_{\Omega \phi}^{J=1 / 2}(k)+\frac{1}{3} C_{\Omega \phi}^{J=3 / 2}(k)+\frac{1}{2} C_{\Omega \phi}^{J=5 / 2}(k).
\label{average}
\end{align}

\section{Results and discussions}
\label{3}

In this section, we present and discuss the numerical results for the $\Omega$--$s\bar{s}$ meson systems.
We first perform bound-state calculations and analyze the low-energy scattering properties in different spin-parity channels.
On this basis, the corresponding scattering parameters are extracted and the dynamical features of the interactions are clarified.
Finally, the femtoscopic correlation functions are evaluated to explore possible experimental signatures and to provide guidance for future measurements.

\subsection{Bound-state calculation and scattering process}
\label{31}

In this part, we investigate the possible bound states and low-energy scattering behaviors of the $\Omega \phi$ and $\Omega \eta^\prime$ systems within the framework of the QDCSM.
Calculations are performed for different spin-parity quantum numbers, taking into account both single-channel and coupled-channel effects.
The resulting energies are summarized in Table~\ref{energy}.
For the $J^{P}=1/2^{-}$ channel, only the $\Omega \phi$ configuration is allowed.
The calculated energy is slightly higher than the corresponding threshold, indicating that no bound state is formed in this channel and that the interaction is weak.

In the $J^{P}=3/2^{-}$ sector, both the $\Omega \eta^\prime$ and $\Omega \phi$ channels contribute.
The single-channel calculations show that neither channel supports a bound state.
When the coupled-channel coupling between $\Omega \eta^\prime$ and $\Omega \phi$ is included, it is insufficient to generate a bound state.
For the $J^{P}=5/2^{-}$ channel, only the $\Omega \phi$ configuration contributes.
The calculated energy is again found above the threshold, implying that the interaction in this channel is not strong enough to support a bound state.
It should be noted that for unbound systems, the present calculations are carried out in a finite model space with a limited number of basis functions.
As the model space is enlarged, the eigenenergy of an unbound state is expected to gradually approach the corresponding threshold from above.

\begin{table}
	\caption{Energies of the single-channel and coupled-channel $\Omega$--$s\bar{s}$ systems (in MeV).}
	\begin{tabular}{c c c c}
		\hline \hline
		~~~~$J^{P}$~~~~ & ~~~~Channel~~~~  & ~~~~$E_\text{th}$~~~~ & ~~~~$E$~~~~  \\ \hline
		$1/2^{-}$  & $\Omega \phi $        & 2662 & 2666   \\
		$3/2^{-}$  & $\Omega \eta^\prime $ & 2554 & 2560       \\
		           & $\Omega \phi $        & 2662 & 2668          \\
		           & $\Omega \eta^\prime $--$\Omega \phi $ & 2554 & 2559   \\
		$5/2^{-}$  & $\Omega \phi $         & 2662 & 2669     \\
		\hline \hline
		\label{energy}
	\end{tabular}
\end{table}

\begin{figure}[htb]
	\centering
	\includegraphics[width=8cm]{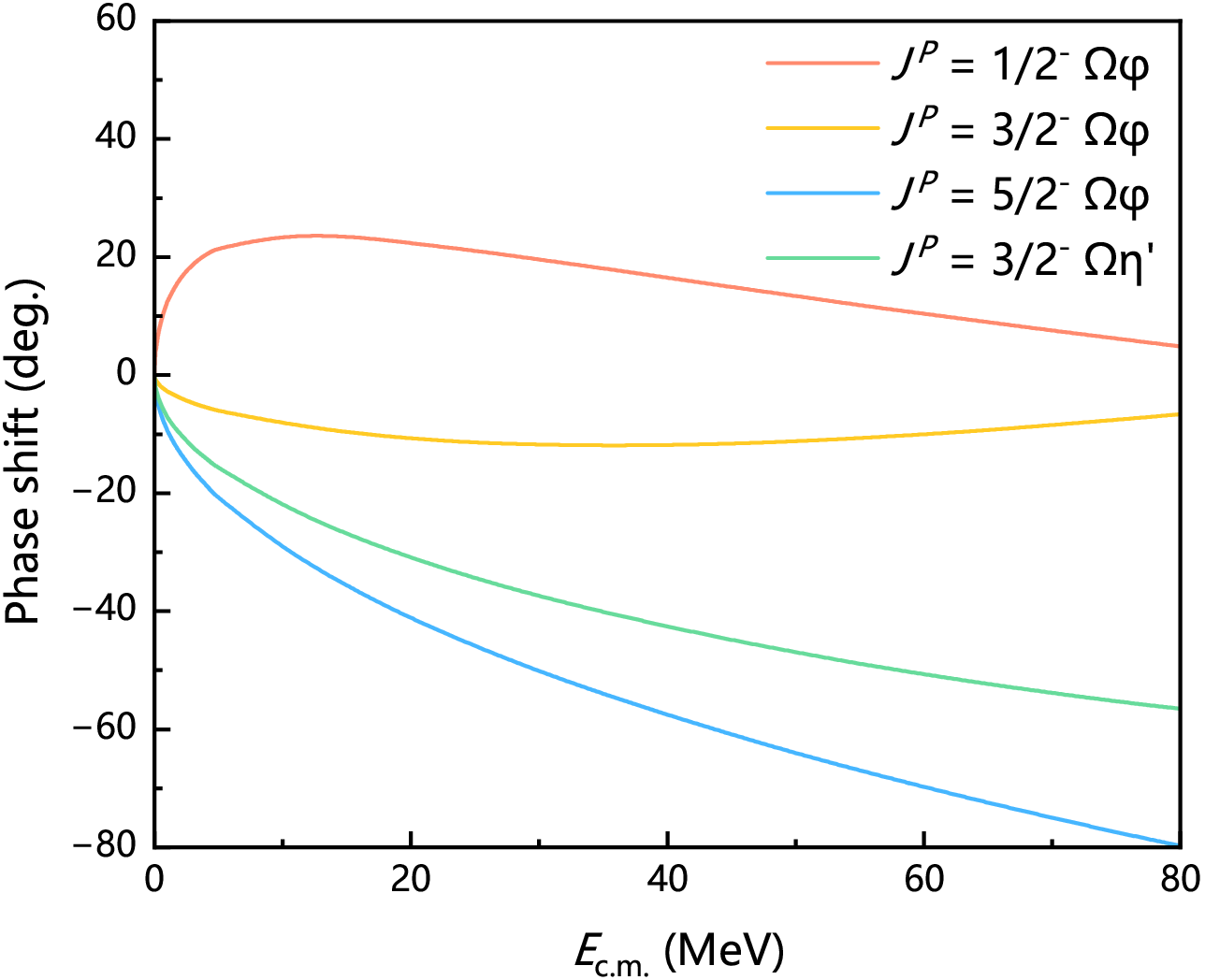}\
	\caption{Scattering phase shifts of the $\Omega$--$s\bar{s}$ systems.}
	\label{scattering}
\end{figure}

To further clarify the nature of the interaction in each channel, we next calculate the $S$-wave scattering phase shifts for the $\Omega\phi$ and $\Omega\eta^\prime$ systems, as shown in Fig.~\ref{scattering}.
For the $\Omega\phi$ system with $J^{P}=1/2^{-}$, the phase shift is positive at low energies and decreases gradually with increasing center-of-mass energy, indicating a weakly attractive interaction.
This attraction, however, is not strong enough to produce a bound state, in agreement with the energy spectrum analysis discussed above.

In contrast, the $\Omega\phi$ channels with $J^{P}=3/2^{-}$ and $5/2^{-}$ exhibit negative phase shifts over the entire energy region considered, with magnitudes that increase monotonically as the energy increases.
Such behavior is characteristic of repulsive interactions and explains the absence of bound states in these channels.
A similar trend is observed for the $\Omega\eta^\prime$ system with $J^{P}=3/2^{-}$, whose phase shift remains negative, indicating a predominantly repulsive interaction.
Although the coupled-channel dynamics between $\Omega\eta^\prime$ and $\Omega\phi$ slightly enhances the attraction, it remains insufficient to generate a bound state.
Overall, the phase-shift analysis provides a consistent and complementary description of the interaction mechanisms in different spin-parity channels, supporting the conclusions drawn from the bound-state calculations.

Based on the calculated low-energy scattering phase shifts, the corresponding scattering length and effective range are extracted by fitting the phase shifts to the effective-range expansion,
\begin{align}
	k \cot\delta = -\frac{1}{a_{0}} + \frac{1}{2} r_{\text{eff}} k^{2} + \mathcal{O}(k^{4}),
\end{align}
which is valid in the low-momentum region.
The obtained low-energy scattering parameters are summarized in Table~\ref{scatteringparam}.
These quantities provide a quantitative characterization of the interaction strength and its range in different spin-parity channels.

\begin{table}
	\caption{Scattering length and effective range for the $\Omega$--$s\bar{s}$ systems.}
	\begin{tabular}{c c c}
		\hline \hline
		~~~$J^{P}$~~~ & ~Scattering length~  & ~Effective range~  \\ \hline
		$\Omega\phi$ with $1/2^{-}$  & $-$1.29 fm & 1.53 fm  \\
		$\Omega\phi$ with $3/2^{-}$  & 0.26 fm & 0.80 fm  \\
		$\Omega\phi$ with $5/2^{-}$  & 0.91 fm & 0.70 fm  \\
		$\Omega\eta^\prime$ with $3/2^{-}$  & 0.71 fm & 1.15 fm  \\
		\hline \hline
		\label{scatteringparam}
	\end{tabular}
\end{table}

We can find that for the $\Omega\phi$ system, the strength of the interaction exhibits a clear spin dependence: channels with lower total spin experience stronger attraction, whereas those with higher spin are characterized by increasingly repulsive interactions.
On the one hand, this behavior is analogous to the well-known $N\Lambda$ system, in which the spin-singlet interaction is more attractive than the spin-triplet one~\cite{Mihaylov:2023ahn}.
On the other hand, such a spin dependence can be naturally understood at the quark level in terms of the Pauli exclusion principle.
In the $\Omega\phi$ system, four identical $s$ quarks are present. In the $S$-wave configuration, the spatial wave functions of these quarks are symmetric, while the color degree of freedom is limited to the three available colors $(r,g,b)$.
For higher-spin states, the four $s$ quarks are more likely to have their spins aligned, which enhances the symmetry of the spin wave function.
This configuration is increasingly constrained by the Pauli exclusion principle and therefore leads to a stronger effective repulsion between the clusters.

\subsection{Femtoscopic correlation function}
\label{32}

\begin{figure}[htb]
	\centering
	\includegraphics[width=8cm]{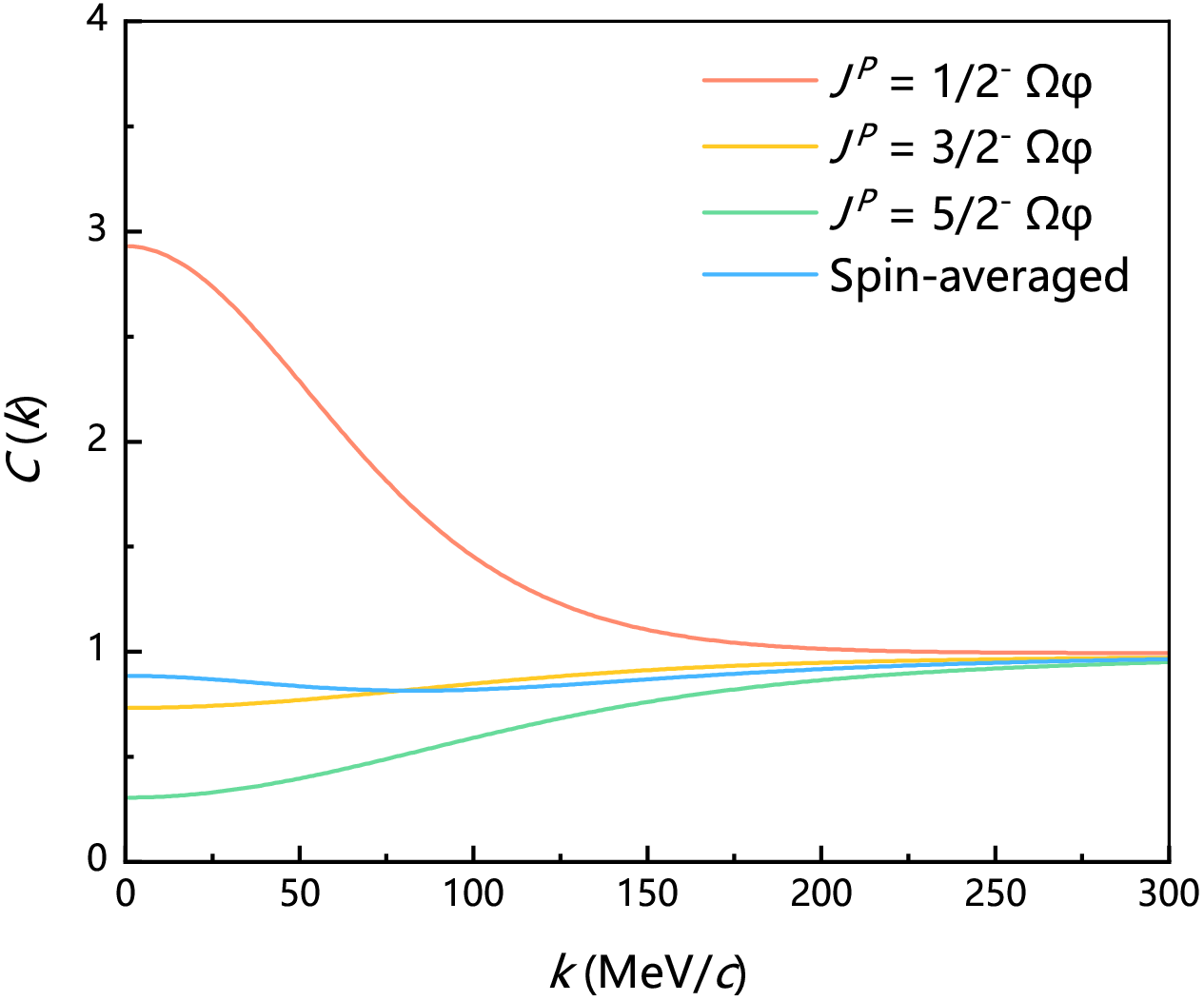}\
	\caption{$\Omega \phi$ correlation functions of different spin channels with source size $R = 1$ fm.}
	\label{CF1}
\end{figure}

To connect the microscopic interaction obtained from the quark model with experimentally accessible observables, we investigate the femtoscopic correlation functions of the $\Omega$–$s\bar{s}$ systems for the next step.
Femtoscopic correlations provide a powerful tool to probe hadron–hadron interactions at low relative momentum, especially for systems involving unstable particles, where direct scattering experiments are not feasible.
Based on the interaction information extracted from the scattering analyses, we calculate the correlation functions to further clarify the nature and strength of the $\Omega \phi$ and $\Omega \eta^\prime$ interactions, and to assess their possible experimental signatures.
Based on the LL model introduced in Sec.~\ref{23}, the femtoscopic correlation functions are calculated using the scattering parameters extracted from the numerical analysis of the scattering processes presented in the previous part.
Fig~\ref{CF1} shows the correlation functions $C(k)$ for different spin-parity channels, as well as the spin-averaged result.

In Fig~\ref{CF1}, the source size parameter is fixed to $R = 1$ fm, which is a typical value adopted in femtoscopic analyses of baryon–baryon and baryon–meson systems in high-energy collisions.
Additionally, recent investigations of the emission source characteristics have been reported in Refs.~\cite{ALICE:2020ibs,ALICE:2023sjd,Mihaylov:2023pyl,VazquezDoce:2024nye,Wang:2025htd}.
As shown in Fig~\ref{CF1}, the correlation functions exhibit a strong spin dependence at small relative momentum.
For the $J^{P}=1/2^{-}$ $\Omega\phi$ channel, a pronounced enhancement of $C(k)$ is observed in the low-$k$ region, reflecting the strong attractive interaction inferred from the positive scattering phase shift and the corresponding negative scattering length.
In contrast, the $J^{P}=3/2^{-}$ channel shows only a weak deviation from unity, indicating a relatively mild interaction.
For the $J^{P}=5/2^{-}$ channel, the correlation function is suppressed below unity at small $k$, which is characteristic of a repulsive interaction and is consistent with the negative phase shift behavior.

Since experimental measurements do not distinguish spin quantum numbers, the observed $\Omega\phi$ correlation function corresponds to a spin-averaged quantity. 
As given in Eq.~(\ref{average}), it is constructed as a weighted sum over different spin channels according to their spin degeneracies.
Although the $J^{P}=1/2^{-}$ channel exhibits strong attraction, its effect is partly offset by the repulsive contributions from higher-spin channels, leading to only a moderate enhancement of the spin-averaged correlation function at low relative momentum. 
This highlights the crucial role of spin averaging in realistic comparisons with experimental data.

\begin{figure}[htb]
	\centering
	\includegraphics[width=8cm]{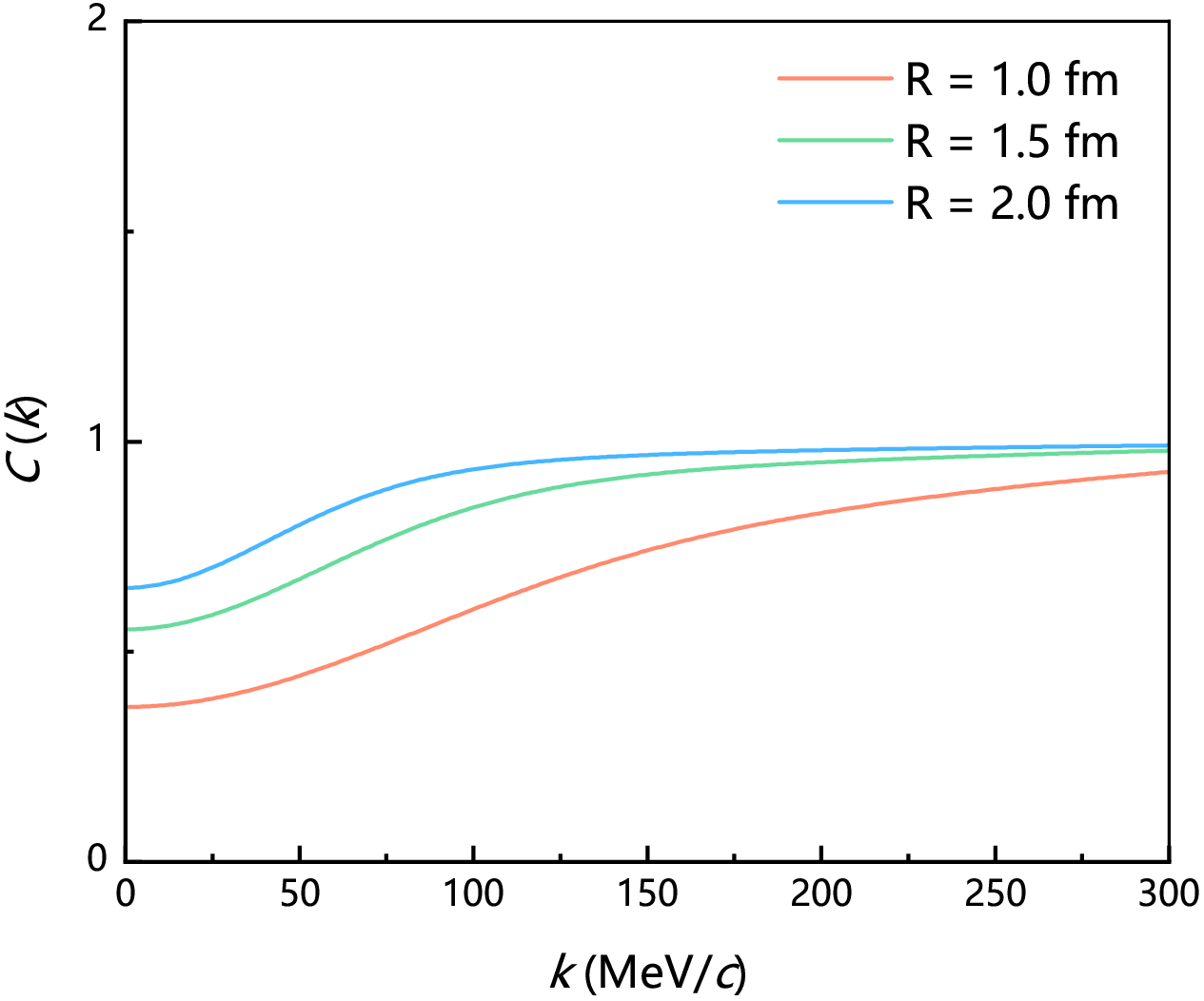}\
	\caption{$\Omega \phi$ correlation functions with different values of the source size parameter $R$.}
	\label{CF2}
\end{figure}

To examine the sensitivity of the $\Omega\phi$ correlation function to the source size, we further calculate the spin-averaged correlation functions with different values of source size $R$, as shown in Fig~\ref{CF2}.
As can be seen from the figure, the $\Omega\phi$ correlation function exhibits only a mild dependence on the source size in the considered momentum region. For smaller source sizes, a slightly stronger suppression of the correlation function is observed at low relative momentum. 
With increasing source size, the correlation function gradually approaches unity, and the low-$k$ structure becomes less pronounced. 
This behavior reflects the fact that a larger emission source effectively smears out the sensitivity to short-range strong interactions, reducing their impact on the correlation signal.

It is worth noting that the overall variation of the correlation function with respect to source size $R$ remains moderate, indicating that the $\Omega\phi$ system is not dominated by a strong near-threshold attraction or bound-state effect. 
This feature is consistent with the scattering analysis presented in the previous subsection, where no bound states are found and the interaction strength is shown to be channel dependent and partly repulsive after spin averaging.

Moreover, the weak source-size dependence is closely related to the spin-averaged nature of the observable. 
As discussed above, the strong attraction in the $J^{P}=1/2^{-}$ channel is largely compensated by repulsive contributions from higher-spin channels. Consequently, the net interaction entering the correlation function is relatively weak, making the resulting correlation less sensitive to the precise value of the emission source size. 
This characteristic may pose challenges for experimentally extracting detailed information on the $\Omega\phi$ interaction, but at the same time provides a clean testing ground for theoretical models once high-precision femtoscopic data become available.

\begin{figure}[htb]
	\centering
	\includegraphics[width=8cm]{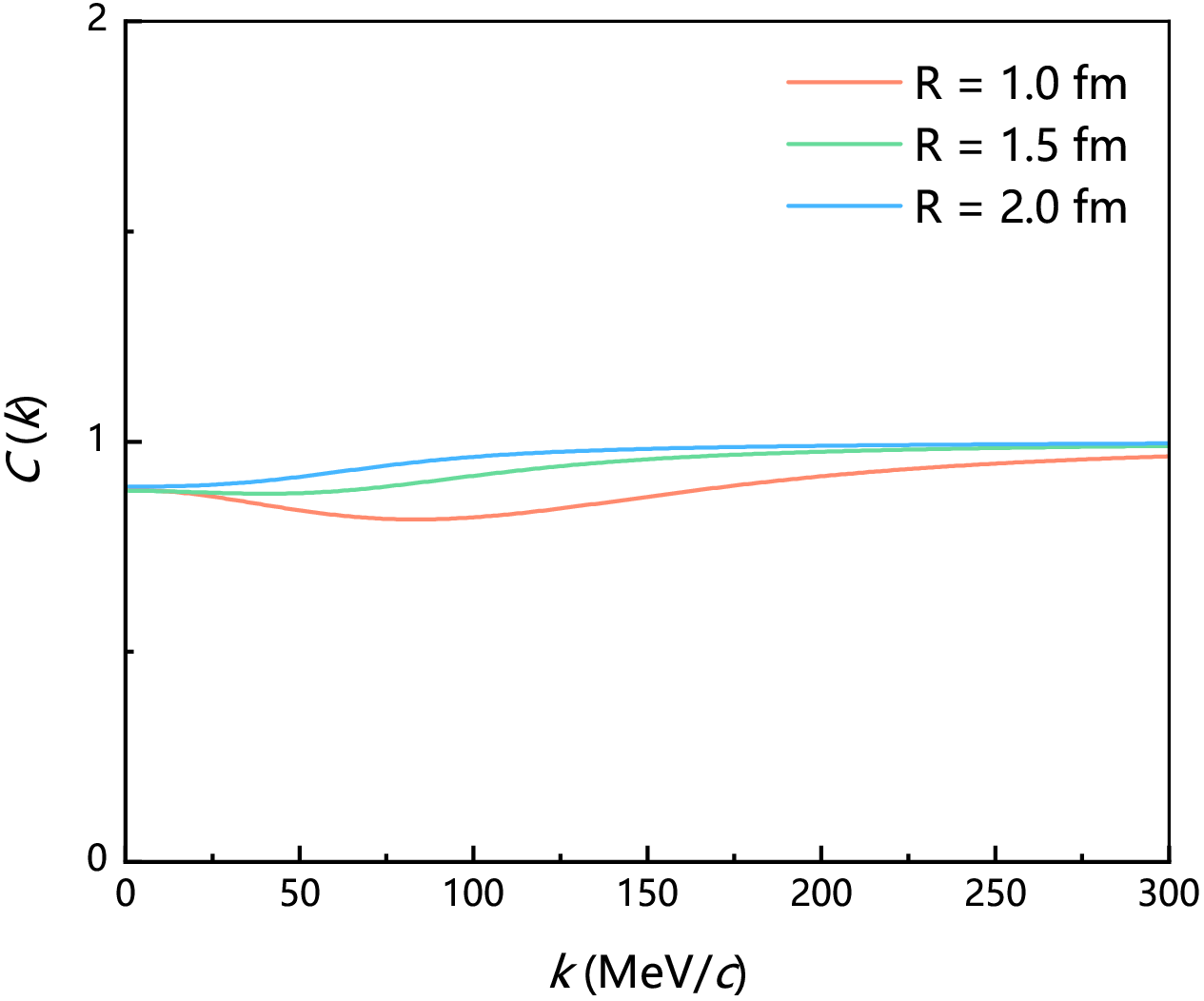}\
	\caption{$\Omega \eta^\prime$ correlation functions with different values of the source size parameter $R$.}
	\label{CF3}
\end{figure}

Fig~\ref{CF3} shows the spin-averaged femtoscopic correlation functions of the $\Omega\eta^{\prime}$ system calculated with different values of the source size parameter $R$. 
The correlation function exhibits a clear dependence on the emission source size, especially in the low relative momentum region.
For small source sizes, the correlation function shows a pronounced suppression at low $k$, indicating that the $\Omega\eta^{\prime}$ interaction in the relevant spin channel is predominantly repulsive. 
As the source size increases, the low-momentum suppression becomes gradually weaker, and the correlation function approaches unity over the whole momentum range. This trend reflects the reduced sensitivity of larger sources to short-range repulsive interactions.

Compared with the $\Omega\phi$ system, the source-size dependence of the $\Omega\eta^{\prime}$ correlation function is slightly more pronounced. 
This behavior can be traced back to the scattering parameters extracted in the previous subsection, where the $\Omega\eta^{\prime}$ interaction in the $J^{P}=3/2^{-}$ channel is characterized by a positive scattering length and a relatively moderate effective range, consistent with a repulsive or weakly interacting scattering state. 
As a result, the correlation function is mainly governed by the suppression induced by short-range repulsion, which is more sensitive to the spatial extent of the emission source.

It is also worth emphasizing that, since only a single spin-parity channel contributes to the $S$-wave $\Omega\eta^{\prime}$ system, the correlation function is not affected by spin-averaging effects, in contrast to the $\Omega\phi$ case. 
This makes the $\Omega\eta^{\prime}$ correlation function a relatively cleaner probe of the underlying two-body interaction, although its overall deviation from unity remains modest.
These features suggest that future femtoscopic measurements of the $\Omega\eta^{\prime}$ system, combined with a careful control of the source size, could provide complementary constraints on the $\Omega$-meson interaction, especially in the absence of bound-state signals.

\section{Summary}
\label{4}

In this work, we have carried out a systematic investigation of the interactions between the $\Omega$ baryon and the $s\bar{s}$ mesons within the framework of the QDCSM. 
Focusing on the $S$-wave $\Omega\phi$ and $\Omega\eta^{\prime}$ systems with different spin-parity quantum numbers, we studied their energy spectrum, low-energy scattering behavior, and corresponding femtoscopic correlation functions.
For the $\Omega\phi$ system, the calculated energies in the $J^{P}=1/2^{-},3/2^{-}$, and $5/2^{-}$ channels all lie above their respective thresholds, indicating unbound scattering states. 
Similar conclusions are obtained for the $\Omega\eta^{\prime}$ system.

To further clarify the nature of the interactions, we analyzed the scattering phase shifts and extracted the low-energy scattering parameters. 
The phase shift analysis reveals that the $\Omega\phi$ interaction is attractive in the $J^{P}=1/2^{-}$ channel, while it becomes repulsive in the higher-spin channels. 
For the $\Omega\eta^{\prime}$ system, the interaction in the $J^{P}=3/2^{-}$ channel is predominantly repulsive. 
The observed spin dependence of the interaction can be understood in terms of quark-level dynamics, where the Pauli principle among the four identical $s$ quarks plays an important role, especially in higher-spin configurations.

The results based on the LL model show that, due to the spin-averaging effect, the repulsive contributions from higher-spin $\Omega\phi$ channels partially compensate the attraction in the $J^{P}=1/2^{-}$ channel, resulting in a moderate spin-averaged correlation signal. 
In addition, the total $\Omega\phi$ correlation function shows only a weak dependence on the source size.
For the $\Omega\eta^{\prime}$ system, the correlation function exhibits a suppression at low momentum, consistent with the repulsive nature of the interaction. 

Overall, the present study provides a comprehensive quark-model description of the $\Omega$–meson interactions and their femtoscopic signatures. 
Our results offer useful theoretical input for future experimental measurements of $\Omega\phi$ and $\Omega\eta^{\prime}$ correlation functions in relativistic heavy-ion and hadronic collisions, and may help to further elucidate the dynamics of multi-strange hadron interactions.

\acknowledgments{This work is supported partly by the National Natural Science Foundation of China under Contracts Nos. 12205249, 12305087 and 12575088. 	
Y. Y. is supported by the Scientific Research Foundation of Changzhou College of Information Technology under Grant No. SG210201B13003.  
Y. W is supported by the Funding for School-Level Research Projects of Yancheng Institute of Technology under Grant No. xjr2025010.
Y. T. is supported by the Qinglan Project of Jiangsu Province of China.
Q. H. is supported by the Start-up Funds of Nanjing Normal University under Grant No. 184080H201B20.
}	

\setcounter{equation}{0}
\renewcommand\theequation{A\arabic{equation}}

\end{document}